\documentclass[sigchi]{acmart}

\AtBeginDocument{%
  \providecommand\BibTeX{{%
    \normalfont B\kern-0.5em{\scshape i\kern-0.25em b}\kern-0.8em\TeX}}}

\copyrightyear{2024}
\acmYear{2024}
\setcopyright{rightsretained}
\acmConference[CHI '24]{Proceedings of the CHI Conference on Human Factors in Computing Systems}{May 11--16, 2024}{Honolulu, HI, USA}
\acmBooktitle{Proceedings of the CHI Conference on Human Factors in Computing Systems (CHI '24), May 11--16, 2024, Honolulu, HI, USA}
\acmDOI{10.1145/3613904.3642777}
\acmISBN{979-8-4007-0330-0/24/05}

\usepackage{xspace}

\begin{document}

\title{AXNav: Replaying Accessibility Tests from Natural Language}


\author{Maryam Taeb}
\email{mr21cg@fsu.edu}
\authornote{Work done while Maryam Taeb was an intern at Apple}
\affiliation{%
  \institution{Florida State University}
  \country{USA}
}

\author{Amanda Swearngin}
\email{aswearngin@apple.com}
\affiliation{
    \institution{Apple} 
    \country{USA}
}

\author{Eldon Schoop}
\email{eldon@apple.com}
\affiliation{
    \institution{Apple} 
    \country{USA}
}

\author{Ruijia Cheng}
\email{rcheng23@apple.com}
\affiliation{
    \institution{Apple} 
    \country{USA}
}

\author{Yue Jiang}
\authornote{Work done while Yue Jiang was an intern at Apple}
\email{yue.jiang@aalto.fi}
\affiliation{
    \institution{Aalto University} 
    \country{Finland}
}

\author{Jeffrey Nichols}
\email{jwnichols@apple.com}
\affiliation{
    \institution{Apple} 
    \country{USA}
}

\renewcommand{\shortauthors}{Taeb, Swearngin, Schoop, et. al.}
\newcommand{\systemname}{AXNav\xspace}

\begin{abstract}
Developers and quality assurance testers often rely on manual testing to test accessibility features throughout the product lifecycle. Unfortunately, manual testing can be tedious, often has an overwhelming scope, and can be difficult to schedule amongst other development milestones.
Recently, Large Language Models (LLMs) have been used for a variety of tasks including automation of UIs. However, to our knowledge, no one has yet explored the use of LLMs in controlling assistive technologies for the purposes of supporting accessibility testing.
In this paper, we explore the requirements of a natural language based accessibility testing workflow, starting with a formative study. From this we build a system that takes a manual accessibility test instruction in natural language (e.g., ``Search for a show in VoiceOver'') as input and uses an LLM combined with pixel-based UI Understanding models to execute the test and produce a chaptered, navigable video. In each video, to help QA testers, we apply heuristics to detect and flag accessibility issues (e.g., Text size not increasing with Large Text enabled, VoiceOver navigation loops). We evaluate this system through a 10-participant user study with accessibility QA professionals who indicated that the tool would be very useful in their current work and performed tests similarly to how they would manually test the features. The study also reveals insights for future work on using LLMs for accessibility testing. 
\end{abstract}



\begin{teaserfigure}
  \includegraphics[width=\textwidth]{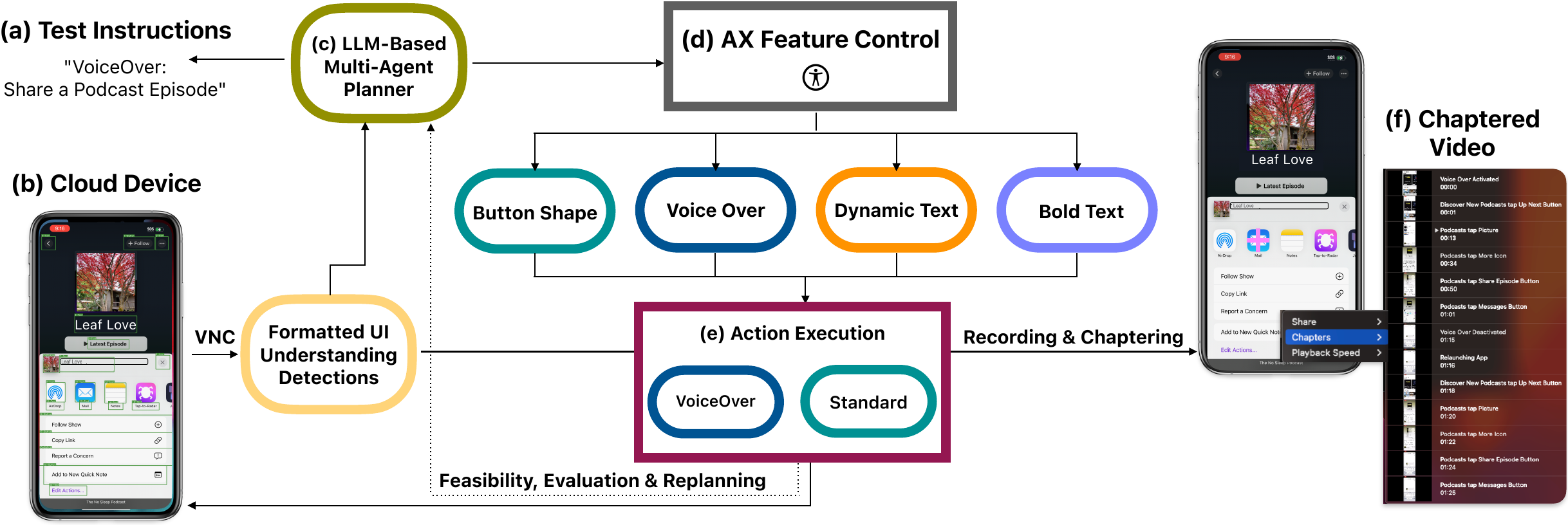}
  \caption{\systemname interprets accessibility test instructions specified in natural language, executes them on a remote cloud device using an LLM-based multiagent planner, and produces a chaptered video of the test annotated with heuristics that highlight potential accessibility issues.
  To execute a test, \systemname provisions a cloud iOS device; stages the device by installing the target app to be tested and enabling a specified assistive feature; synthesizes a tentative step-by-step plan to execute the test from the test instructions; executes each step of the plan, updating the plan as needed; and annotates a screen recording of the test with chapter markers and visual elements that point out potential accessibility issues.
  }
  \Description{The figure shows a flow diagram where on the left, a Cloud Device and Test instructions are shown. The middle shows a diagram which visualizes the workflow described in the caption. On the far right, the diagram shows a mobile device with a Chapters menu open on the right showing the steps marked by the system.}
  \label{fig:teaser}
\end{teaserfigure}

\received{14 September 2023}
\received[revised]{12 December 2024}
\received[accepted]{19 January 2024}

\maketitle

\section{Introduction}
Many mobile apps still have incomplete support for accessibility features \cite{fok2022large, ross2017epidemiology, yan2019current, alshayban2020accessibility, zhang2021screen}. 
Developers of these apps may not implement or test accessibility support due to a lack of awareness \cite{alshayban2020accessibility}, organizational support \cite{ross2017epidemiology,  bi2022accessibility}, or experience in accessibility testing \cite{bi2022accessibility}.
For apps that do support accessibility features, developers often work in tandem with experienced accessibility quality assurance (QA) testers~\cite{bi2022accessibility}. Employees in both roles may use automated tools like accessibility scanners~\cite{accessibilityInspector,accessibilityScanner}, linters~\cite{androidLint}, and test automation~\cite{espresso, xctest} to execute UI test scenarios.
However, despite many available tools, the majority of testing for accessibility is still done manually. This may in part be due to the limitations of the tools themselves. For instance, UI tests can be brittle~\cite{pan2020gui,li2017atom} or non-existent~\cite{lin2020test, kochhar2015understanding, cruz2019attention}, and scanners can provide false positives~\cite{vigo2013benchmarking}. In addition, manual testing can reveal issues that cannot be detected by automated techniques alone~\cite{mankoff2005your}.

However, manually testing all possible accessibility scenarios and features is costly and hard to scale.
In a formative study with six accessibility QA testers, we found they often had difficulties keeping up with the scope of apps and features they were assigned to test. This causes testers to limit the scope of their tests, potentially letting bugs slip through, and can lead to test instructions becoming outdated.
While research has addressed some of these challenges through automation~\cite{salehnamadi2021latte, salehnamadi2023assistive}, there are still manual costs associated with writing and recording tests to be replayed. Recorded tests often need to be updated when the UI or navigation flow changes, similar to UI automation tests, which must specify each step in the navigation flow in code~\cite{li2017atom, pan2020gui}. 

To address some of these challenges and support existing manual testing workflows of accessibility QA testers, we explore the use of \textit{natural language instructions} to specify accessibility testing steps to a system. Manual test instructions are common artifacts within organizations that often have large databases of manual steps for QA testers.
Our system, \systemname, interprets natural language test instructions to produce a set of concrete actions that can be taken in an app, which it then adapts automatically as the interface evolves. \systemname executes these actions on a live cloud device, enabling and configuring accessibility features as needed, and runs heuristics on target screens to flag potential issues to manual testers. \systemname's output is a chaptered, annotated video that captures the interaction trace along with heuristic results.

Our approach is motivated by prior work that uses Large Language Models (LLMs) to recreate bug reports~\cite{feng2023prompting}, test GUIs~\cite{liu2023chatting}, and automate tasks for web interfaces~\cite{shaw2023pixels}. 
To our knowledge, \systemname is the first work that uses LLMs for accessibility testing, or controlling accessibility services~\cite{voiceover} and settings~\cite{buttonshapes}.

The contributions of this work are:
\begin{itemize}
    \item A formative study with 6 professional QA and accessibility testers revealing motivation and design considerations for a system to support accessibility testing through natural language instruction-based manual tests. 
    \item A novel system, \systemname, that converts manual accessibility test instructions into replayable, navigable videos by using a large language model and a pixel-based UI element detection model. The system helps testers pinpoint potential issues (e.g., non-increasing text, loops) with multiple types of accessibility features (e.g., Dynamic Text, VoiceOver) and replays tasks through accessibility services to enable testers to visualize and hear the task as a user of the accessibility service might perform it. 
    \item A user study with 10 professional QA and accessibility testers revealing key insights into how accessibility testers might use natural language-based automation within their manual testing workflow. 
\end{itemize}

\section{Related Work}
\systemname is most closely related to works that use text instructions as an input for UI automation, which is useful beyond accessibility use cases. In this work, we specifically target UI navigation from natural language for accessibility testing, thus we also review accessibility testing tools and approaches. 
\subsection{Large Language Models and UI interaction}

A key contribution of \systemname is its LLM-based planner that can navigate mobile apps to execute specific tasks or arrive at particular views.
Our multi-agent system architecture is loosely based on ResponsibleTA, which presents a framework for facilitating collaboration between LLM agents for web UI navigation tasks~\cite{responsibleTA}. Since \systemname is designed for testing rather than end-user automation, it removes some components (e.g., a system to mask user-specific information), and combines other modules (e.g., \systemname combines evaluation and completeness verification, and \systemname proposes actions and feasibility in the same step). These changes significantly reduce the number of LLM turns taken, which lowers cost and reduces latency.

Other UI navigation works for web and mobile apps have recently emerged. Wang et al. ~\cite{bryanConversationLLM}, describe prompting techniques to adapt LLMs for use with mobile UIs, and evaluate an LLM-based agent's ability to predict the UI element that will perform an action on a given screen. \systemname's UI navigation system builds upon this work by supporting more complex, multi-step tasks.
Other works map from detailed, multi-step instructions to actions in mobile apps~\cite{feng2023prompting, yangMappingGrounded, UGIF}. AutoDroid injects known interaction traces from random app crawls into an LLM prompt to help execute actions with an LLM agent~\cite{wen2023empowering}.
\systemname can interpret a wide variety of instruction types, from highly specific step-by-step instructions to unconstrained goals within an app (``add an item to the cart''), without relying on prior app knowledge. Furthermore, \systemname is able to modify its plan when the UI changes, if it encounters errors, or if the test instructions are incorrect.

The emergence of LLM-based UI navigation systems has motivated the need for more interaction datasets. Android in the Wild presents a large dataset of human demonstrations of tasks on mobile apps for evaluating LLM-based agents~\cite{androiditw}. Other datasets, such as PixelHelp~\cite{yangMappingGrounded} and MoTiF~\cite{burns2022motif} also collect mobile app instructions and steps.
Unlike prior art, \systemname is designed to work on iOS apps, which can have different navigation flows and complexities than corresponding Android apps.

Most importantly, none of the above works have been used to interact with accessibility features or support accessibility testing workflows. This is the core focus of \systemname's contribution.

\subsection{Accessibility Testing Tools}
Despite the availability of accessibility guidelines and checklists \cite{accessibilityApple,accessibilityGoogle, wcag}, linters and scanners \cite{accessibilityScanner, accessibilityInspector, androidLint}, and platforms for test automation \cite{xctest, espresso}, developers and QA testers still often prefer to test their apps manually~\cite{linares2017developers, lin2020test}. Testing manually by using accessibility services can reveal issues that cannot be revealed by scanners alone \cite{mankoff2005your}. However, manual testing is costly and difficult to scale, leading to a variety of automated tools and testing frameworks being developed for accessibility testing~\cite{kochhar2015understanding}. 

There are a variety of tools to automatically check accessibility properties of apps~\cite{silva2018survey}. Development-time~\cite{androidLint} approaches use static analysis to examine code for potential issues. Run-time tools~\cite{accessibilityScanner,espresso,robolectric2021,accessibilityInspector} examine a running app to detect accessibility issues, which enables them to detect issues beyond static analysis; however, they still must be activated on each screen of the app to be tested. 

Another approach is to automatically crawl the app to detect issues~\cite{eler2018automated, chen2021empirical, alshayban2020accessibility, salehnamadi2022groundhog}; however, such tools currently adopt random exploration and thus may not fully cover or operate the UI as an end-user might. 
These crawlers also do not operate through accessibility services which leaves them unable to evaluate whether navigation paths through the app are fully accessible. 

Latte \cite{salehnamadi2021latte} starts to bridge this gap by converting GUI tests for navigation flows into accessibility tests that operate using an accessibility service; however, the majority of apps still lack GUI tests \cite{lin2020test} and often require updating the code to new navigation flows when a UI changes~\cite{pan2020gui}. Removing the requirements for GUI tests to be available, A11yPuppetry \cite{salehnamadi2023assistive} lets developers record UI flows through their app and replay them using accessibility services (i.e., TalkBack~\cite{talkback}). This idea has also been explored in prior work for web applications~\cite{bigham2010accessibility}.
However, a key challenge with record and replay approaches is that they can also be brittle and difficult to maintain as the UI evolves~\cite{pan2020gui,li2017atom}.
By using LLMs, \systemname can interpret plain text instructions at different levels of granularity, and adapt them to new context when UIs change.

\systemname was not intended to fully scan apps for accessibility issues. Rather, it was designed to flag a subset of potential issues during test replay to aid manual accessibility QA testers, based on feedback from formative interviews.
Our system architecture could also be extended to run accessibility audits during each step of the replay, similar to accessibility app crawlers~\cite{salehnamadi2022groundhog, eler2018automated}; however, in this work we focus on navigation and replay through accessibility services and not on holistic reporting of accessibility issues.

\begin{figure}
    \centering
\includegraphics[width=\columnwidth]{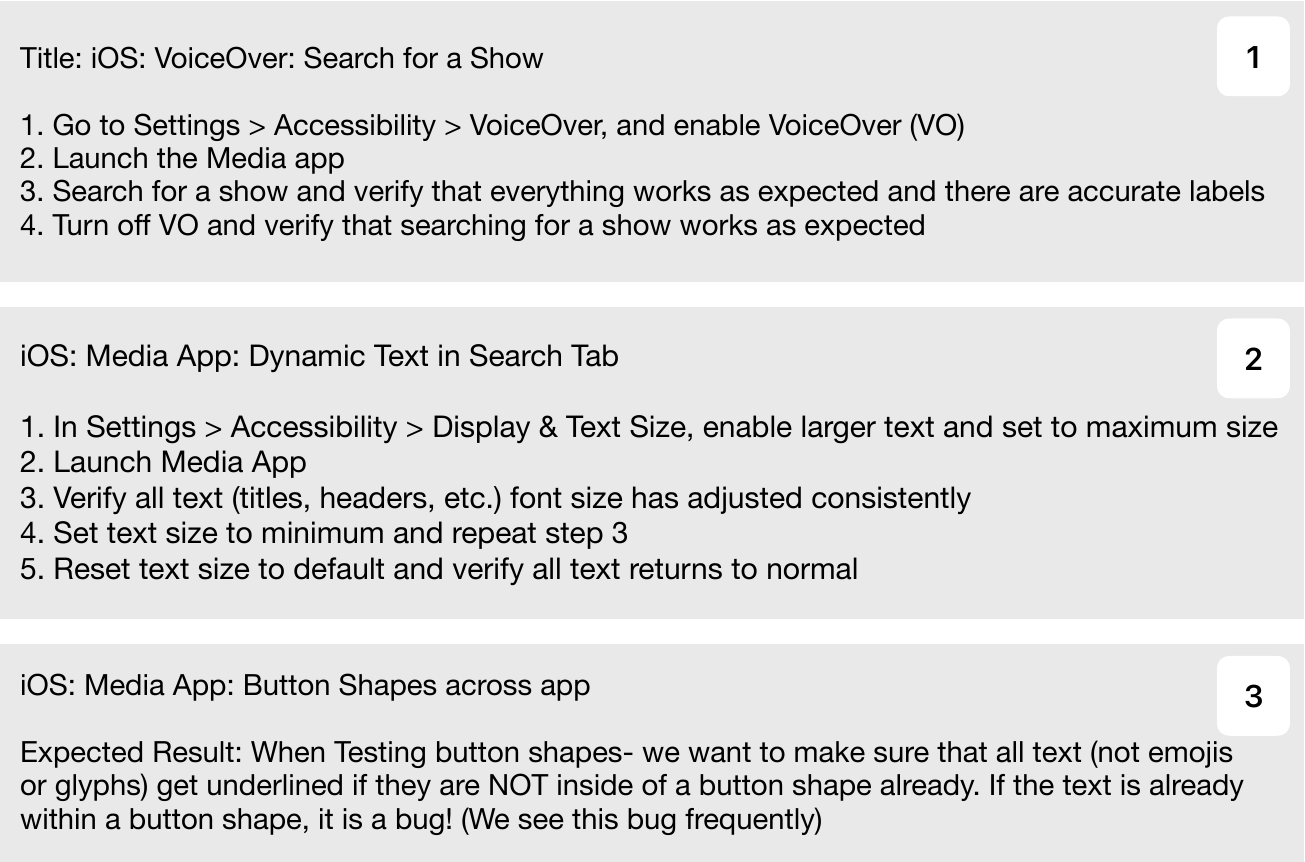}
    \caption{Three sample test cases for a video streaming media app testing the accessibility features of VoiceOver, Dynamic Type, and Button Shapes. Testing instructions typically consist of a title containing the app and feature under test, and a set of manual test instructions in natural language. The tests may also contain expected result descriptions. Some tests have specific, low-level instructions (1,2) and others give only a high-level instruction (3).}
    \Description{
        The figure shows three test cases from our dataset. 
    }
    \label{fig:test_examples}
\end{figure}

\section{Formative Interviews}
\label{section:formative_interviews}
To better understand the challenges and benefits of manual accessibility testing and elicit requirements for \systemname, we recruited six accessibility QA professionals through snowball recruiting at a large technology company. Participants spanned four product and services teams across four organizations, and had a minimum of 3 years of professional experience in accessibility and QA testing of iOS mobile apps. We conducted 30-minute remote interviews with each participant. 

We divided our formative study into two parts. In the first part, we asked participants about the challenges and benefits of manual accessibility testing, their cadence for performing manual tests, and whether and how they write testing instructions. We also asked them to describe the areas and features they tested and to demonstrate a manual test for an app and feature of their choice. 

From our domain knowledge and review of prior work, we hypothesized that a significant portion of time spent testing was manually navigating to specific screens in apps, and that a system to automatically perform this navigation from existing manual test instructions would be useful. The second half of the formative study was designed to check this assumption and elicit features that would be useful for a system to help support manual testing.
In this phase, we played a screen recording of an author manually performing an accessibility test from an internal database of existing tests (Figure~\ref{fig:test_examples}.1---``Search for a Show'' in a media app using VoiceOver). We asked participants to imagine a system replaying the test instructions on the device and instructed them to think aloud while watching the screen recording, noting any features an automated tool should support. We asked about the benefits and drawbacks of this functionality and how it might be used in testing workflows, if at all. We include the full set of formative interview questions in our supplementary materials.

\subsection{Challenges \& Benefits of Manual Testing}
Participants noted a key benefit of manual testing is to experience the feature as an end user might (P2-P5). One participant, P3, being a VoiceOver user, mentioned this enables them to more realistically test the feature as it is meant to be used: ``the advantages are that we can test literally, from the user perspective myself, and a number of my teammates are users of the features because of various accessibility needs that we have. So we are the foremost experts in the functionality of those particular tests and what the expected results would be.'' (P3) 

Participants from three teams brought up challenges, including an overwhelming scope of features and scenarios, leaving them to target only a few key features and tasks for testing (P2, P4-P6). P5 stated: ``It's not like necessarily difficult. It is just like, repetitive and kind of boring and the scope is so big a lot of times like, if you're looking at the \textit{<App Name Anonymized>} app,
there's so many pages and so many views and so many buttons and different types of elements and everything. That is overwhelming and you feel you're going to miss something''.

Writing manual tests was also noted as a challenge by participants from three teams, who write down or have existing test suites of manual instructions (P3-P6). Participants noted it was easy for those tests to become outdated when apps are updated, challenging less experienced QA testers' ability to interpret and follow test instructions (P3, P4). Finding the right time for accessibility testing was also mentioned by four participants, as they worked with apps that are frequently updated across various product milestones (P3-P6). Two participants also mentioned trying to develop automated tests in their work, which they described as easily breaking and not covering all possible scenarios (P5, P6).  

\subsection{Testing Process}
All participants took part in accessibility testing at various times throughout the product lifecycle. They tested annually as new features were added, or on regular release cycles of app interfaces.  The participants' daily work consists of manually performing tests for accessibility features (e.g., VoiceOver, Dynamic Type) across various products, or additionally writing accessibility frameworks and automation code. 

To test purely visual accessibility features, the participants typically toggle on the feature under test and validate that the app's UI renders or behaves correctly based on the setting. For accessibility services tests (e.g., VoiceOver), they typically enable the feature, and then either navigate the app to perform a task using the feature or navigate to a specific screen to validate the navigation order or another behavior of the feature.

\subsection{Granularity and Availability of Manual Accessibility Test Instructions}
Manual testing instructions are an extremely common artifact within our organization, existing in both manual test databases and bug-tracking tools. One team we interviewed (two participants) noted they own a large database of manual instructions for UI tests (P3, P4), but none of these instructions are specifically for accessibility testing. They also noted that they frequently write down manual instructions for accessibility features, or ``repro steps'', when they are filing bugs. In their work, they often work with engineers who may lack familiarity with the accessibility feature under test, so they try to make instructions as specific as possible.

For another two participants on a different team, their accessibility testing instructions primarily consisted of a large regression test suite across ten apps with 300 individual test cases they perform annually (P5, P6). Among these tests, some had concrete low-level steps, but many were abstract, high-level, and assumed the QA tester has a high level of expertise on both the app and the accessibility feature to be tested.
\autoref{fig:test_examples} contains three example test cases for a video streaming app for the accessibility features Voice Over, Dynamic Type, and Button Shapes. Each test case typically has a title containing the platform, feature, and app to be tested, but only some test cases have step by step instructions, and only some test cases have an ``Expected Result'' specified. 


\subsection{Features in a Natural Language-Based Accessibility Testing Tool}
In the second part of our formative study, we elicited features by having participants imagine a system replaying manual testing instructions on an iPhone, while watching a screen recording of one of the authors performing a manual test. The video was a screen recording only and had no additional features. We then asked participants what features such a system should support in the context of accessibility testing. Here we summarize the key features revealed by both this task and part one of our interviews that we incorporated into the design of \systemname.

\subsubsection{F0: Natural Language Interpretation and Replay}
\label{subsection:f0}
Our QA testers liked to observe the behavior of the interactions as they were performing manual testing. They wished for more automation in their workflows, but did not have time to spend writing and updating automated tests. They also often already had large databases of manual testing instructions available. Thus one goal of our work was to \textit{enable testers to use their existing testing instructions, written at multiple levels of abstraction, as input to a system that can interpret those instructions and replay them on a device}. We hypothesized such a system could complement testers' workflows through automation without requiring writing and updating fully automated tests.

\subsubsection{F1: Quickly Navigate and Visualize Executed Steps}
\label{subsection:f1}
To provide QA testers with the benefit of observing tests as an end user, we record videos of each test for the tester to examine. While watching the video demonstration of the test, multiple participants requested to review portions of the video multiple times to better understand what action the system took and to further examine screens for potential bugs. To improve video navigation, we add chapter labels to the video that indicate either the action taken or flag potential issues. We also annotate system actions on the impacted video frames with a pink `+' cursor. The chapters also allow users to skip back and repeat watching key segments quickly (Figure~\ref{fig:teaser}.f).

We also received feedback from two participants during our interviews requesting the system to let them replay the instructions on a live local device and take control during various parts of the test (P3, P4). This would be a more useful interaction particularly for P3, a VoiceOver user, as they were unable to interact with the UI in the video format. Due to current constraints with our system architecture, we did not provide this in \systemname, but will explore the feasibility of supporting this along with providing a video for post-replay review.

\subsubsection{F2: Flag Potential Issues}
\label{subsection:f2}
Four participants mentioned that they would like the system to flag potential issues and report failures. When asked what specific issues to flag would be most helpful, participants mentioned both visual issues like dynamic type resizing, and accessibility feature navigation issues (e.g., wrong navigation order, elements missing a label or not available for navigation).
Participants noted that if a system could direct them to target their testing towards any potential issues, that would save time in bug filing: ``If it could detect the issue and write it down or like, ..., that would be helpful so that I can write bugs or maybe bug can be automated.'' (P6) Based on this feedback, we developed custom heuristics in \systemname to flag a small subset of accessibility issues to evaluate the feasibility and potential impact of this idea. We use the video output to flag issues by adding a chapter label at the location of the potential issue in the video. 

Some participants also requested the system to save screenshots in addition to the video output (F1) so that when they find issues, they can directly upload the screenshots to a bug tracking tool -- ``if your product did that, I think that would be a huge time saver because most of my time is taking screenshots and clipping'' (P5). Screenshots also enable \systemname to flag potential visual accessibility issues through postprocessing. 

\subsubsection{F3: Realistic VoiceOver Navigation and Captioning}
\label{subsection:f3}
In the video recording of the manual test, we showed participants, we activated the UI elements for each step directly like a sighted user might, rather than swiping through elements on a screen to find UI elements as a non-sighted user might. Several participants noticed this, and noted that the system should \emph{replay the test to be as similar as possible to how a user of the accessibility feature performs the task} (P2, P3-P5). Additionally, our video also included the VoiceOver \textit{captions panel} for this task, which three participants mentioned was an important feature to include in our final system. 

\subsubsection{F4: Perform Tests With and Without Accessibility Features} 
\label{subsection:f4}
Our participants shared many manual test scripts that instructed testers to perform tests with and without the accessibility feature under test toggled on. For example, the ``Search for a Show'' test in Figure~\ref{fig:test_examples} instructs the tester to first turn on VoiceOver to perform the test, and to perform the same test after turning off VoiceOver. As participants noted, testing with the feature turned on and off helps QA testers verify if the system returns to the correct state after turning off the feature under test. Thus, \systemname repeats the navigation steps twice for most tests, first replaying the test with the feature on and then replaying the test with the feature off. 
\section{\systemname System}
Based on our formative interviews with QA testers, we designed and built \systemname, a system that interprets an accessibility test authored in natural language, and replays the test instructions on a mobile device while manipulating the accessibility feature to be tested (F0; \autoref{subsection:f0}).
\systemname interprets plain text instructions, which can be authored at varying levels of specificity, to navigate to a desired view to be tested. It then outputs a chaptered video that a tester can navigate and replay (F1; \autoref{subsection:f1}) annotated with heuristics that flag potential issues in the app (F2; \autoref{subsection:f2}).
\systemname currently supports controlling and flagging issues with four accessibility features: VoiceOver, a gesture-based screen reader~\cite{voiceover}; Dynamic Type, which increases text size; Bold Text, which increases text weight; and Button Shapes, which ensures clickable elements are distinguishable without color, typically by adding an underline or button background (Figure~\ref{fig:teaser}.d). We selected these features since, based on our interviews, they seemed to provide good coverage of real-world testing needs across different modalities. \systemname could be extended to other accessibility and device features in the future.
For each user-provided test, \systemname executes the test on a specified app both with and without the specified assistive feature activated for comparison (F4; \autoref{subsection:f4}).

\systemname consists of three main components that are used to prepare for, execute, and export test results: (1) Device Allocation and Control, (2) Test Planning and Execution, and (3) Test Results Export. These components work together to provision and stage a cloud iOS device for testing, automatically navigate through an app running on the cloud device to execute the test, and collect and process test results.

\begin{figure*}[!h]
    \centering
    \includegraphics[width=\textwidth]{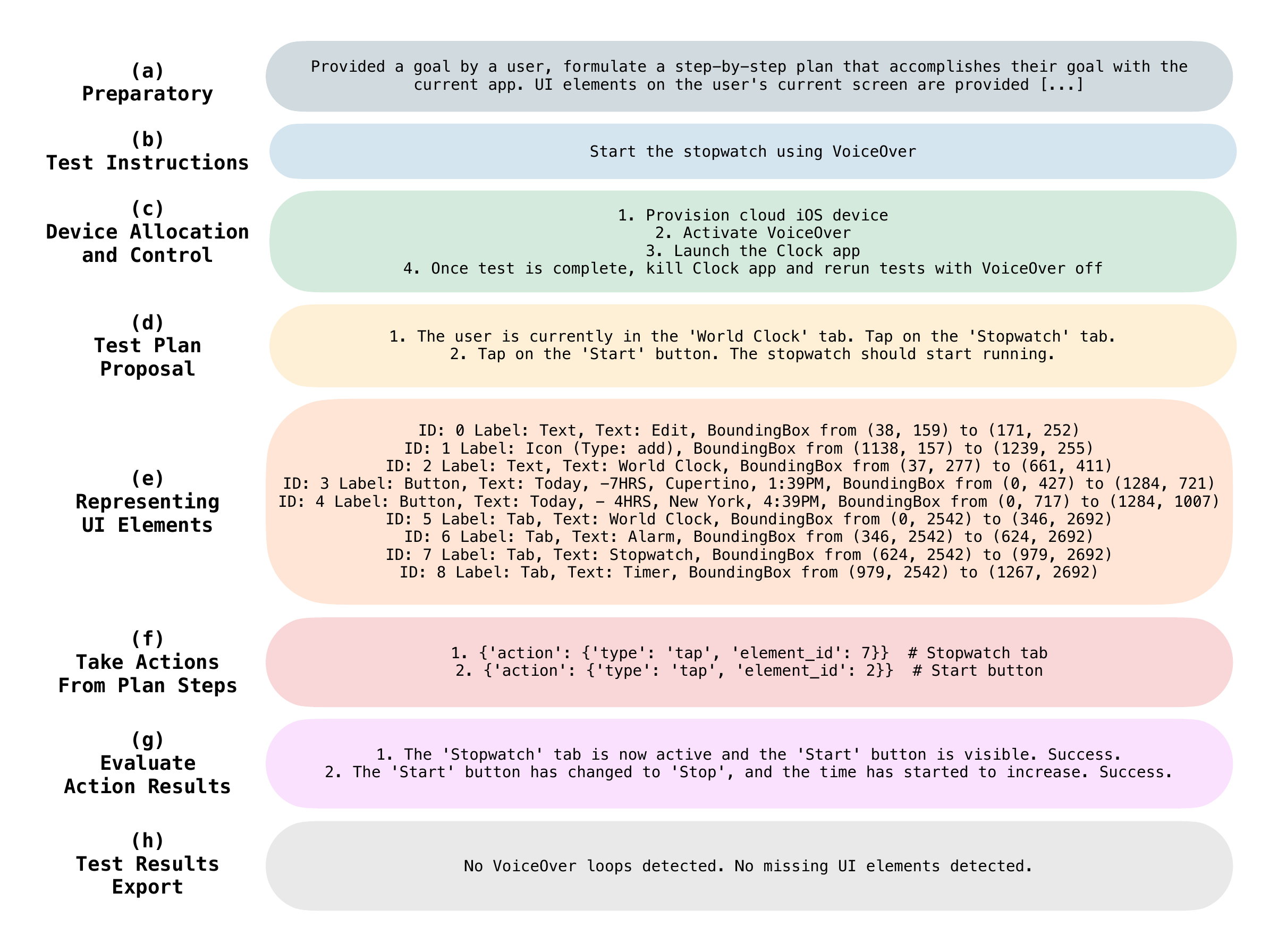}
    \caption{Overview of intermediate steps used by \systemname to interpret natural language test instructions; provision and stage a device for testing; formulate and execute a plan to navigate the UI for the test; and export the test results.}
    \Description{The figure shows the different phases of our system in separate rows, including Preparatory, which contains the high-level goal and prompt; Test instructions includes the high-level test instruction; Device Allocation and Control includes the steps the system takes to initialize the device; Test Plan Proposal shows the high-level test plan the planner produced; Representing UI Elements shows formatted UI detections for an example screen; Take Actions from Plan Steps shows examples of the concrete predicted actions to take on the UI; Evaluate Action Results shows examples of planner evaluation results; Finally, Test Results Export shows the output of the test, which is used to generate the output video.}
    \label{fig:llm-framework}
\end{figure*}

\subsection{Device Allocation and Control}
\label{sec:system.setup}
Before executing a test, \systemname provisions a remote cloud iOS device and prepares it according to the parameters it extracts from the test instructions. \systemname extracts the name of the app to be tested and the assistive technology to use in the test (e.g., Dynamic Type) from the instructions to automatically install the app and select the assistive feature to test. Instructions typically take the form of those shown in Figure~\ref{fig:test_examples}. 

During setup, \systemname installs a custom application that provides an interface to operating system APIs that silences several system notifications, controls screen recording, and interacts with assistive technologies.
\systemname uses an operating system API to toggle and configure the specific accessibility feature under test (e.g., Dynamic Type size). If the test is for VoiceOver, \systemname activates the caption panel (F3; \autoref{subsection:f3}) and sets the speaking rate to 0.25 to accommodate for speeding up the exported video in the Test Results Export step.

When the device is ready for the test to be executed, \systemname launches the app under test, and begins screen recording. The test execution engine can interact with the cloud device over a remote desktop connection and the accessibility-specific features supported by the custom application (see \autoref{sec:system.voautomate}).

\subsubsection{Accessibility Feature Control and Replay}
\systemname uses different sequences to test supported accessibility features.
For tests with Dynamic Type, the system launches the target application, increases the Dynamic Type Size, navigates to the target screen specified in the test, takes a screenshot, kills the application, and repeats this process for all four Dynamic Type sizes and, finally, without Dynamic Type on. This enables testers to observe the corresponding changes on the screen as the size is increased.

For Bold Text and Button Shapes, \systemname navigates to the target screen specified in the test with and without the feature enabled, and saves pairwise screenshots of each tested screen with the feature on and off for comparison. 

For VoiceOver, \systemname replays the instructions once with VoiceOver toggled on, and again with VoiceOver off.

\begin{figure}
    \centering
\includegraphics[width=\columnwidth]{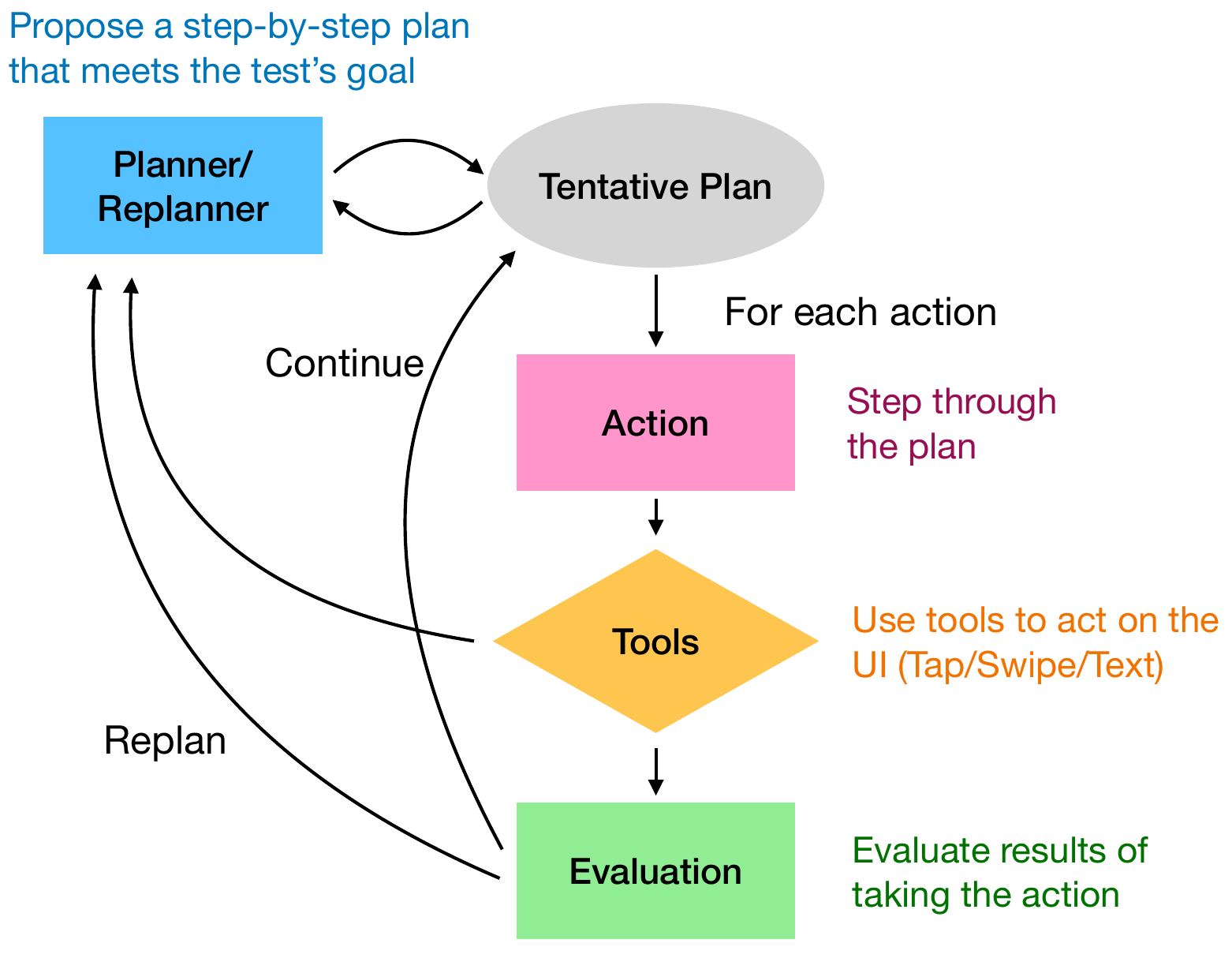}
    \caption{Planning and replanning workflow of our LLM-Based Multi-Agent Planner}
    \Description{
        The figure shows the planning and replanning workflow of our system. A Planner produces a Tentative Plan which directs to Action which steps through the plan. Action directs to Tools which uses tools to act on the U (e.g., Tap, Swipe), and Action directs to Evaluation which evaluates the results of taking an action. Evaluation points to Replan which goes back to Planner. Evaluation can also direct to Continue which goes back to Tentative to continue the flow. 
    }
    \label{fig:llm_ui_nav}
\end{figure}

\subsection{Test Planning and Execution}

\systemname uses an LLM-based UI navigation system that can translate from natural language test instructions into a set of actionable steps, execute steps on a live device by calling APIs that interact with a device, and feed results back to improve the navigation plan (see \autoref{fig:llm_ui_nav}).
We use OpenAI GPT-4~\cite{gpt4} in our implementation, but \systemname can be easily adapted to use other LLMs.
Our system architecture is loosely inspired by ResponsibleTA~\cite{responsibleTA}, but eliminates some elements (e.g., masking LLM inputs), and merges other elements (e.g., combining feasibility with actions). It consists of three LLM-based agents: the planner agent, the action agent, and the evaluation agent.
To provide device state to the LLM agents, we use existing pixel-based machine learning models to recognize UI elements,  text, and icons~\cite{zhang2021screen, CompleteIcon}. \systemname formats detected UI elements as text strings to be ingested by the LLM, described in \autoref{sec:system.planning}.
To interact with the device, \systemname provides tools that the LLM invokes to send touch or keyboard input events and VoiceOver gestures.

\subsubsection{Test Plan Proposal}
\label{sec:system.planning}
The planner agent is the heart of \systemname (\autoref{fig:llm_ui_nav}), and it formulates a tentative plan containing instructions to navigate to a desired view in an application from its current state. The planner agent takes as input the accessibility test instructions (\autoref{fig:llm-framework}.b), the name of the app under test, and the formatted UI element detections from a screenshot of an iOS device. The planner agent's prompt contains instructions to formulate a tentative plan (\autoref{fig:llm-framework}.a; \autoref{fig:llm_ui_nav}, Tentative Plan) to accomplish the test goal with the current app and the set of actions that can be taken in a step.
To adapt to changes in the UI or unexpected errors (e.g., permissions request dialogs), the prompt includes instructions to traverse backward through the app if an unexpected state is encountered, and to accept an imperfect plan if needed, since it can be revised later. The planner agent's prompt also instructs the model to to provide reasonable search queries if the test does not specify them, based on the app name and the current context of the screen. 

The expected output of the planner agent is a JSON-formatted object that contains a list of steps. Each step contains a \texttt{thought} designed to facilitate Chain-of-Thought (CoT) reasoning~\cite{letsThinkStepByStep} that answers how the step will help achieve the user's goal; \texttt{evaluation}, which suggests criteria to determine task success; \texttt{action}, a brief, specific description of an input to provide on a given screen (e.g., tap, swipe, enter text); and a \texttt{status} field, which is initialized as ``todo'' and updated to ``success'' when a step is executed correctly. An illustrative plan is shown in \autoref{fig:llm-framework}.d.

\subsubsection{Representing UI Elements to the Agents}
\systemname describes the UI to the LLM as a list of UI elements in plain text, which each contains an incrementing integer as an id; the classification of the UI element (e.g., Icon, Toggle); text contained by the UI element, if any; and the coordinates of the bounding box around the element (\autoref{fig:llm-framework}.e). For example, an element with ID 3 might appear as:
\texttt{(3) [Button (Clickable)] "Try It Free" (194, 1563) to (1042, 1744)}. \systemname uses this simplified list because it economizes on tokens, unlike prior approaches that format UI elements as JSON or HTML~\cite{responsibleTA, feng2023prompting}. 

\systemname infers the elements in a UI using the Screen Recognition model from Zhang et al.~\cite{zhang2021screen} to predict bounding boxes, labels, text content, and the clickability of UI elements from screenshot pixels of iOS devices.
Using pixels to detect UI elements makes \systemname agnostic to the underlying UI framework~\cite{prefab}.
\systemname groups and sorts detected elements in reading order, and flags an element if it is recognized as a top-left back button, using the postprocessing approaches from~\cite{zhang2021screen}.
\systemname also detects the presence of a keyboard (to hint that a text field is selected) by detecting the presence of single-character OCR results on the lower third of a screenshot. If \systemname detects a keyboard, it filters all UI elements detected on the keyboard, except for a submit button (``return'', ``search'', ``go'', etc.).



\subsubsection{Mapping from Plan Steps to Concrete Actions}
For each step in the plan proposed by the planning prompt, \systemname implements an LLM-based ``action agent'' to map from the text instruction to a concrete action (Figure~\ref{fig:llm_ui_nav}, Action) to take on a particular UI element (\autoref{fig:llm-framework}.f), inspired by prior work~\cite{yangMappingGrounded, feng2023prompting}. This agent performs several critical subtasks to navigate UIs in a single step: it identifies how to map a natural language instruction to the specific context of a UI, evaluates the feasibility of the requested action, and produces arguments for a function call to execute the task.
The action agent's subtask-to-action prompt contains instructions to output a specific action to take on a given screen, represented by the formatted UI detections. The available actions are:


\begin{itemize}
\item \texttt{Tap}: Tap a UI element given its ID. The prompt instructs the agent that tapping an object that is inferred to be non-clickable is acceptable if it is the only reasonable option on a screen.
\item \texttt{Swipe}: Swipe in a cardinal direction (up/down/left/right) from a specified (x, y) coordinate. The system tells that agent that swiping can be used to scroll to view more options available on a screen if needed.
\item \texttt{TextEntry}: Tap a UI element given its ID and then enter a given text string by emulating keystrokes. The agent is told to come up with appropriate text if it is not provided.
\item \texttt{Stop}: Stop execution of the current step and prepare feedback for the replanner to update the plan as needed. The feedback must specify what information is needed in an updated plan.
\end{itemize}

The output of the action agent is a JSON-formatted object that contains a \texttt{thought} to elicit CoT reasoning, \texttt{relevant UI IDs}, a list of UI elements the agent considers relevant (also to elicit CoT reasoning), and a single \texttt{action}, which specifies a function call in JSON to execute interactions on the device.

\subsubsection{Executing VoiceOver Actions}
\label{sec:system.voautomate}
For action execution in VoiceOver, the system interacts with the device through VoiceOver's accessibility service. We implement this in a custom application that provides an interface to a Swift API (built on top of XCTest~\cite{xctest}) that can trigger key VoiceOver gestures~\cite{voiceover} for \systemname.
These gestures execute VoiceOver gestures in the same way a user of VoiceOver would perform them (F3; \autoref{subsection:f3}).
Supported gestures are as follows: 

\paragraph{Right swipe through all elements \texttt{(read-all)}}
This command triggers the VoiceOver \textit{Right Swipe} gesture multiple times to navigate through all exposed elements on the screen, typically in a top-left to bottom-right ordering. Our system limits the number of elements navigated to 50 to save time and avoid getting stuck in loops or screens with infinite scroll. After right-swiping through the first 50 elements, the system activates the first tab, if it exists, and navigates through all tabs from left to right in the tab bar. 

\paragraph{Activate an element \texttt{(activate-from-coordinates)}}
This command issues VoiceOver's \textit{Right Swipe} and \textit{Double Tap} gestures to locate and activate an on-screen element.
In our formative interviews, our prototype video demonstrated the ``Search for a Show'' task in VoiceOver by directly navigating to relevant UI elements using the VoiceOver \textit{Tap} gesture followed by \textit{Double Tap}.
However, participants gave us feedback that they preferred the demonstration to be more similar to how a non-sighted user would find and activate a UI element, by using \textit{Right Swipe} to navigate through UI elements to find the target UI element to activate, and then activating the element using \textit{Double Tap} (F3; \autoref{subsection:f3}).
To confirm this, we observed a screen reader user performing the ``Search for a Show'' task, who followed a roughly similar pattern.

\texttt{activate-from-coordinates} takes as input \texttt{x} and \texttt{y} coordinates corresponding to the center of the UI detection bounding box to be activated; and the \texttt{UI Type} label from the UI detection model (e.g., \texttt{Tab}). If the \texttt{UI Type} is \texttt{Tab}, the system navigates the VoiceOver cursor directly to the leftmost tab element, uses the \textit{Right Swipe} gesture to swipe to the first tab containing the \texttt{x} and \texttt{y} coordinates, and then activates it using \textit{Double Tap}.
If the \texttt{UI Type} is not \texttt{Tab}, the system navigates forward from the current element using \textit{Right Swipe} until it reaches the last VoiceOver element or finds an element containing \texttt{x} and \texttt{y} which it activates using \textit{Double Tap}. If the system does not find the element, it navigates backward using \textit{Left Swipe} until it reaches the first VoiceOver element containing \texttt{x} and \texttt{y} and if so, activates it using \textit{Double Tap}. If the system does not find an element containing the coordinates, the command returns without activating any element.

\paragraph{Scroll (Up/Down/Left/Right) \texttt{(scroll-<direction>)}}
This command issues the VoiceOver \textit{Three Finger Swipe} gesture, which scrolls the current screen in the given cardinal direction by one page.




To prevent the VoiceOver caption panel from interfering with the UI detection model's assessment of the state of the app, the system removes the caption panel from the formatted UI detections using a heuristic based on a fixed height from device dimensions.
When the input test instructions specify to perform a task that requires navigating through multiple UI elements and screens, the system triggers VoiceOver navigation using \texttt{activate-from-coordinates} when the action agent instructs a \textit{TextEntry} or \textit{Tap} action. If the action agent instructs the system to perform a \texttt{Scroll} action, the system calls the corresponding \texttt{scroll-<direction>} action in VoiceOver.
If the instructions state to navigate to a specific screen to verify the VoiceOver elements and navigation order, the system calls \texttt{read-all} once it reaches the final step of UI navigation, to swipe through all exposed elements on the screen. This enables testers to determine whether all elements within that screen are accessible by VoiceOver.

\subsubsection{Evaluation and Replanning}
\label{sec:system.evalreplan}
Once an action is executed on the device, \systemname implements a third LLM-based ``evaluation agent'' to evaluate the results of the taken action (\autoref{fig:llm_ui_nav}, Evaluation).  An illustrative example of evaluation output is shown in \autoref{fig:llm-framework}.g. 

\systemname prompts the evaluation agent with the test goal, the entire current tentative plan, the action JSON object (including the function call and ``thought''), the UI detections of the screen before the action was taken, and UI detections of the screen after the action was taken. The prompt also includes evaluation hints designed to reduce navigation errors: if UI elements significantly change, the action likely succeeded; if the state of the current screen changes, but a new view is not opened, err on the side of the action succeeding; if the last action was a scroll or swipe, but the screen did not change, the action likely failed; if the target element is not visible, more scrolling may be required; and if the last action was to click on a text field, the evaluation should be whether a keyboard is visible.

The output of the evaluation agent is a JSON object that contains \texttt{evaluation\_criteria}, to encourage CoT reasoning; a \texttt{result} of success, failure, or task completion; and an \texttt{explanation}, which the system feeds back into the Planner to revise the plan if the evaluation fails.

If the evaluation result is positive, then execution proceeds with the action agent being prompted with the next step in the plan. If the evaluation result is negative, the planner agent is prompted to replan, which updates the tentative plan from the current step onwards.
The planner agent's replanning prompt is similar to the initial planning prompt, but includes the previous plan, the current step being executed, and information about the stop condition or evaluation error. The resulting JSON output contains a new tentative plan, revised from the current step onward.



\begin{figure}
    \centering
\includegraphics[width=\columnwidth]{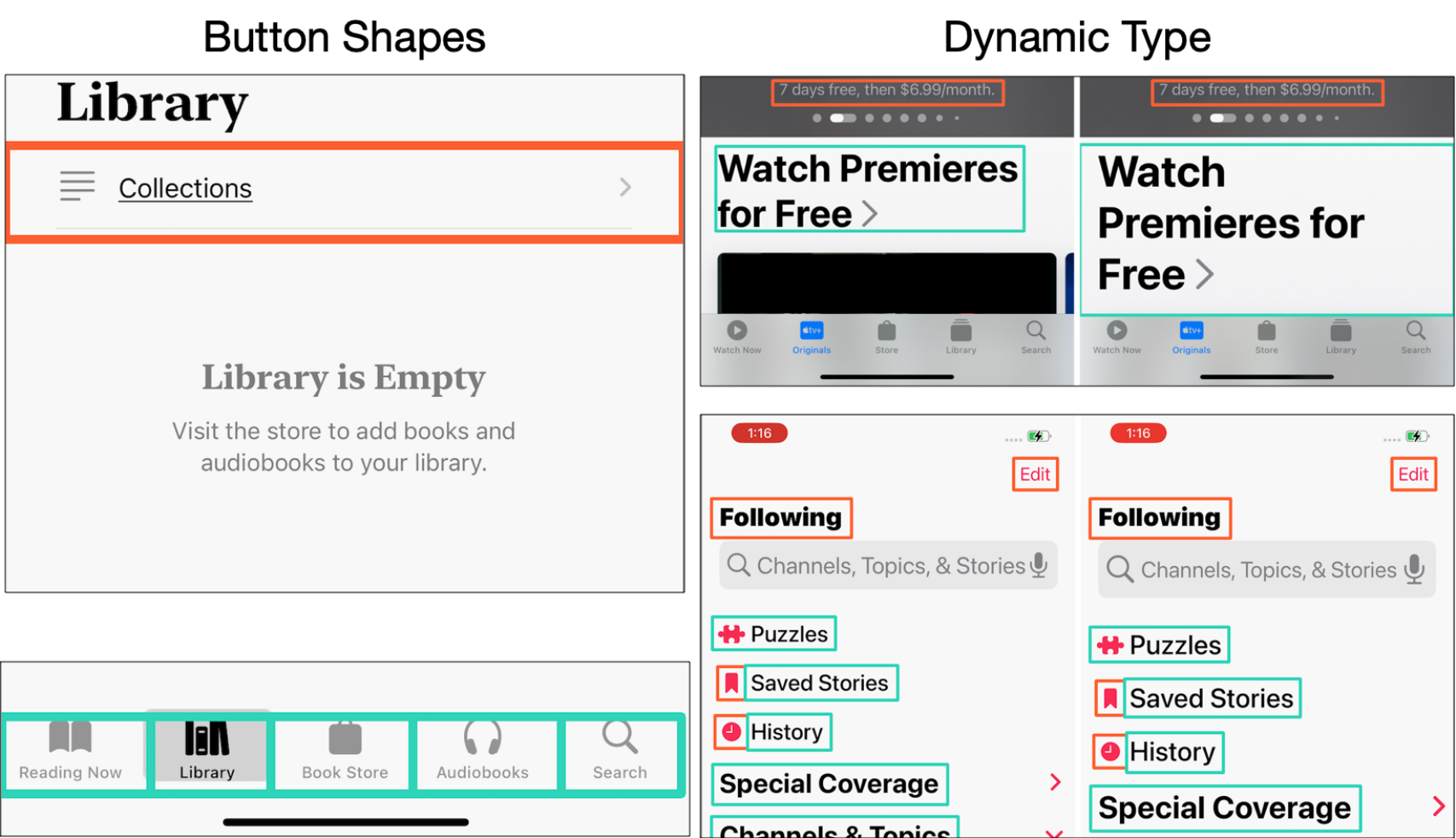}
    \caption{Examples of issues flagged by our heuristics for Button Shapes (left) and Dynamic Text (right). The Button Shapes heuristic flags the Collections row which has a button shape and also is underlined (a possible bug). The Dynamic Type heuristic flags several text elements with red boxes indicating the size has not increased with the DT size update (a possible bug). }
    \Description{The figure shows screenshots of UIs on the left; Button Shapes has a red box flagged around the Collections row which has a Button Shapes bug with an underlined UI element inside a button shape. it also has green boxes around UI elements passing the test; Dynamic Type examples show red boxes around text elements that have not increased in size, and green boxes around text elements that have increased in size between the two screenshots.}
    \label{fig:heuristics}
\end{figure}

\subsection{Test Heuristics}
\systemname can currently flag four types of potential accessibility issues in the output video: VoiceOver navigation loops and missing elements, Dynamic Type text resizing failures, and Button Shapes failures (see \autoref{fig:heuristics}).


\subsubsection{VoiceOver loop detection and missing VoiceOver elements}
Our system detects loops in VoiceOver navigation order during the \texttt{activate-from-coordinates} and \texttt{read-all} commands. To detect loops, the system maintains a list of all visited VoiceOver elements, and detects a looping bug if any element is revisited during the command. To enable the system to navigate the remaining task steps, the system attempts to break out of the loop by finding the next VoiceOver element below the element where the looping was detected, navigating to it, and either continuing with \texttt{read-all} or \texttt{activate-from-coordinates}. 

\subsubsection{Missing VoiceOver elements}
A typical accessibility error occurs when an element detected by \systemname's UI element detection algorithm cannot be navigated by VoiceOver. The system flags this issue when a VoiceOver element cannot be found during \texttt{activate-from-coordinates}. 

\subsubsection{Dynamic Type}
The Dynamic Type heuristic determines if text elements and their associated icons increase in size when the system-wide Dynamic Type size is increased. The heuristic takes two inputs: a screenshot of a view with a baseline text size, and another screenshot of the same view with a Dynamic Type size increased by one increment. 

The heuristic first uses a UI element detection model~\cite{zhang2021screen} on each screenshot to recognize text elements and perform OCR~\cite{appleOCR}. The heuristic then uses fuzzy string matching with Levenshtein distance to find corresponding text elements between the two screenshots, with a partial similarity threshold set to 50\%. The heuristic excludes elements without matches. For a text element to pass the heuristic, its corresponding UI element must increase by an adjustable threshold set to 10\% compared to the baseline screenshot.

To identify icons paired with text elements, which should typically scale along with the text, the heuristic greedily matches icons to text elements in both screenshots by minimizing the distance between the icon's right bounding box coordinate to the text element's left coordinate. To remove icons that are not to the immediate left of the text, the heuristic excludes icons with a gap of more than half the icon's width to the right text element or whose top and bottom are not bounded by the text element's bounding box.
The heuristic pairs icons with their adjacent text elements, and applies the same 10\% threshold in the bounding box area to pass the heuristic. 


\subsubsection{Button Shapes}
The Button Shapes heuristic determines, for a given screenshot, whether clickable text outside of the clickable container is underlined.
This heuristic takes a single screenshot of a view with Button Shapes activated.
The heuristic uses a UI element detection model~\cite{zhang2021screen} to locate and classify elements in the UI, along with their predicted clickability.
For every clickable container element (Buttons and Tabs), the heuristic flags any contained element that is also underlined, which indicates a bug. For any uncontained text element predicted as clickable, the heuristic flags it if it is not underlined. 

The heuristic detects underlines in text elements by extracting the image patch of the text bounding box, binarizing the patch using Otsu's method~\cite{otsuMethod}, edge-detecting the image with the Canny edge detector~\cite{cannyEdge}, and using the Hough Line transform~\cite{houghline} to detect any horizontal line that spans at least 75\% of the width of the patch.
If a text element is underlined when it should not be (or vice versa), it fails the heuristic.

\subsection{Output Video Generation}
\systemname's output is a video of the test execution.
Throughout the replay process, \systemname records the screen of the cloud device and logs timestamps of every action performed on the device, along with actions and activated UI elements.
To improve the navigability of the video, \systemname adds named chapter markers that demarcate each step of the test being performed and each issue flagged by a heuristic (F1 \& F2; \autoref{subsection:f1} \& \autoref{subsection:f2}). Many video players include features to view all chapter markers by name and navigate directly to the start of a given chapter.
To help communicate actions while watching, \systemname overlays markers on the video stream that label each action taken with crosshairs for tap actions and arrows indicating scroll direction. Potential accessibility issues from heuristic results are also overlaid on the video stream with colored bounding boxes in either orange or cyan.
\systemname also speeds up the exported video by a factor of 2.5 to minimize pauses due to the latency of its LLM-based agents.

\section{Technical Evaluation}
\label{section:technical_evaluation}
We conducted two evaluations of \systemname to determine the accuracy of our test replay. Few datasets currently exist in the literature for UI navigation tasks for mobile apps from natural language, and we are aware of no such datasets for iOS apps specifically. Instead, we evaluated the system on a regression test suite used within our company to test a set of media apps, and created our own dataset from free apps within the Apple App Store.

\begin{table}[]
\centering
\begin{tabular}{@{}lllll|llll@{}}
\multicolumn{5}{c}{Regression Testing Apps} & \multicolumn{4}{c}{Performance} \\
\midrule
\textit{Diff.} & \textit{VO} & \textit{BT} & \textit{DT} & \textit{BS} & \textit{Success} & \textit{Partial} & \textit{Fail} & \textit{Acc.} \\
Easy & 17 & 3 & 21 & 3 & 42 & 0 & 2  & \textbf{95.5\%} \\
Hard & 15 & 1 & 2 & 0 & 11 & 2 & 5 & \textbf{61.1\%} \\
\hline
\textbf{Total:} & 32 & 4 & 23 & 3 & \multicolumn{3}{c}{\textbf{Overall Accuracy:}} & \textbf{85.5\%} \\
\end{tabular}
    \caption{Total evaluation test case counts for our Regression Testing Dataset for the AX features of VoiceOver (VO), Dynamic Type (DT), Bold Text (BT), and Button Shapes (BS), which we total for the difficulty level of Easy and Hard respectively. We report the performance of navigation replay as full success, partial success (some but not all steps completed), and failure, along with overall accuracy.}
\label{table:regresson_test_counts}
\end{table}

\begin{table}[]
\centering
\begin{tabular}{@{}lllll|llll@{}}
\multicolumn{5}{c}{Free Apps} & \multicolumn{4}{c}{Performance} \\
\midrule
\textit{Diff.} & \textit{VO} & \textit{BT} & \textit{DT} & \textit{BS} & \textit{Success} & \textit{Partial} & \textit{Fail} & \textit{Acc.} \\
Easy & 0 & 4 & 2 & 1 & 5 & 1 & 0 & \textbf{83.3\%} \\
Hard & 5  & 1 & 3 & 4 & 9 & 3 & 2 & \textbf{64.3\%} \\
\hline
\textbf{Total:} & 5 & 5 & 5 & 5 & \multicolumn{3}{c}{\textbf{Overall Accuracy:}} & \textbf{70.0\%} \\
\end{tabular}
    \caption{Total evaluation test case counts for our Free Apps Dataset for the AX features VoiceOver (VO), Dynamic Type (DT), Bold Text (BT), and Button Shapes (BS), which we total for each difficulty level of Easy and Hard. We report the performance of navigation replay as full success, partial success (some but not all steps completed), and failure, along with overall accuracy.}
\label{table:free_test_counts}
\end{table}

\subsection{Regression Testing Dataset}
First, we evaluated the system on a large regression manual test suite. Some examples of this test suite are shown in Figure~\ref{fig:test_examples}. From that test suite, we extracted 64 test cases from 5 apps testing the accessibility features that \systemname supports: VoiceOver, Dynamic Type, Button Shapes, and Bold Text. We discarded two of the tests due to our account lacking the necessary subscription to view the screen(s) being tested. The final set contains 62 test cases. Note that this regression test suite is used for manual testing and is not constructed for the purpose of being used by any automated system. Many of the tests are very high level and assume the QA tester has a high level of expertise on the feature and the app under test.
We chose to evaluate \systemname on this dataset since it is a representative set of real-world accessibility tests.

\subsection{Free Apps Dataset}
We also constructed a dataset of accessibility testing instructions for publicly available apps. We randomly selected apps from a public list of the 100 most popular free apps in the Apple App Store, ultimately selecting five apps from different app categories. Then for each app, one researcher on our team drafted four manual tests, one for each of \systemname's supported accessibility features, using the regression testing suite as an example. We validated that the tests were realistic by discussing them with an expert accessibility QA tester from the formative study. The final dataset consists of 20 manual tests across five apps and four accessibility features.


\subsection{Accuracy Results}
We evaluated the difficulty of each test through a rubric based on prior work~\cite{gur2023real}, which rates each type of instruction task into Easy or Hard categories for evaluation. 
\begin{itemize}
    \item \textbf{Easy regular expression-based retrieval task}: These tests can be completed in a single step by matching the correct UI element with the correct action, and possibly scrolling on the resulting page. The role of the planner agent in completing these tests is minimal and in many cases, the test could be completed entirely by the action agent.
    \item \textbf{Hard structured problem-solving or open-loop planning task}:  These tests require the system to take multiple actions across multiple screens. That requires the planner agent to reason about the steps needed to complete the test and correct itself as needed as the test proceeds.  It also requires the action and evaluation agents to ensure multiple steps are completed successfully, beyond just the one step required for easy tasks.
\end{itemize}

To group the tests into the above categories, two authors independently rated each test and then met to discuss and resolve any differences. Table~\ref{table:regresson_test_counts} and Table~\ref{table:free_test_counts} show the total counts for each level across the four supported accessibility feature categories and the two separate datasets. 

To repeat each test, we input the test instructions into the system, reset the phone's current state to match the initial state specified by the test, and then executed the test instructions on the device. During this process, we recorded all interactions between the system and the app. For both datasets, we report \textit{navigation replay success}, which measures whether our system can follow the instructed steps successfully to reach the desired destination, and \textit{accessibility test success} for whether the accessibility feature test succeeded. We also report navigation partial success, which indicates that AXNav replayed one or more steps in the test but did not end up in the correct final state. We determined success based on our own manual evaluation based on the expected behavior for each accessibility feature. To ensure consistency, two researchers independently scored the system's performance on each test case and then met to discuss and resolve any differences. 

For the regression testing dataset, our system successfully replayed 95.5\% of easy test cases, and 61.1\% of hard test cases for an overall success rate of 85.5\%. Table~\ref{table:regresson_test_counts} summarizes these results. Within our organization's apps, support for the supported accessibility features is already high; the accessibility test success rate across these tests was 78\%. We are also working with the owners of this regression testing dataset to report the accessibility test failures in our internal bug-tracking system. 

For the free apps dataset, our system successfully replayed 83.3\% of easy test cases, and 64.3\% of hard test cases for an overall success rate of 70.0\%. Table~\ref{table:free_test_counts} summarizes these results. Support for the accessibility features of Bold Text, Dynamic Type, and Button Shapes unfortunately were low across the five apps, resulting in an accessibility test success rate for these apps of only 15.0\% across the 20 test cases. This further motivates the potential impact of using systems like ours within the app development workflow.

While the navigation replay success of our system is good for both datasets, our system fails to replay some tests. In some cases, the navigation replay fails because the test requires tapping on a certain item in a collection where only some items have the required condition (e.g., have a subscription available) but the planner agent typically suggests activating the first item. In other cases, the planner agent cannot deduce enough knowledge about the app and predicts that key functionality for the replay does not exist in the app. In a few cases, key UI elements needed to be activated for the test that were located offscreen and required scrolling to reach, and \systemname did not continue scrolling long enough to find them. Another challenge we have seen is that the planner agent sometimes is unable to determine when to stop and gets into an infinite loop. These are areas we hope to improve in future work.




\section{User Study}
We presented our system in user study sessions with 10 professional accessibility testers. The goal of the user study is to understand how \systemname could assist accessibility testers in their workflows, specifically, how well the system could replicate manual accessibility tests, aid testers in finding accessibility issues, and be integrated into existing test workflow.

\subsection{Procedure}

We conducted 10 1-to-1 interview-based study sessions.  
During each session, we first presented an overview of \systemname to the participant. We then showed three videos generated by \systemname and the associated test instructions, in randomized order.
Each video showed an accessibility test on iOS media applications for e-books, news stories, and podcasts, respectively, with different UI elements and layouts. The videos were selected from the set of videos used in \autoref{section:technical_evaluation}, based on their coverage of different accessibility features, including VoiceOver, Dynamic Type, and Button Shapes. Two of the tests shown in the videos were selected from those with the difficulty level of Easy, and one test with the difficulty level of Hard.
The tests shown in the videos represented real accessibility tests that our participants would perform, as they were selected from the set of test instructions authored and used by testers in the organization.
We chose to show videos to participants as they are the primary output produced by \systemname, offering a realistic representation of interaction with our system. Furthermore, since \systemname is not a production system, it was not optimized for speed, and can take several minutes to an hour to produce a video. In practice, this is not a critical limitation, since many tests can be run in parallel, possibly overnight, and reviewed all at once following their completion.
The specific videos and associated test instructions that we used for the user studies are as follows: 

 \begin{enumerate}
     \item \emph{VO:} This video shows a test of a podcast application. The test instruction prompts the system to share an episode of a podcast show through text message using Voice Over. (Difficulty level Hard)\footnote{This video does not include any issue flagged by the system. In order to show participants what heuristics in VO look like, we presented a supplementary video of another VO case where the system flags a VoiceOver navigation loop in the chapters.}
     \item \emph{DT:} This video shows a test of Dynamic Text in a news application. The test instruction prompts the system to increase the size of the text in four different fonts in a specific tab of the application.  (Difficulty level Easy)
     \item \emph{BS:} This video shows a test of Button Shapes in an e-book application. The test instruction prompts the system to test the Button Shape feature across all the tabs in the application.  (Difficulty level Easy)
 \end{enumerate}

All three videos contained some accessibility issues, which we prompted the participants to discover using the heuristics as part of the system.
Furthermore, all videos deliberately contained errors and imperfect navigation to conservatively showcase the capabilities of our system. Specifically, the VO video shares a podcast \emph{itself} instead of an \emph{episode}, and some false positive errors are flagged in the DT and BS videos.
We intentionally presented those imperfections to the participants to show the performance of the system conservatively, and to trigger a discussion of limitations and future directions. 

For each video, the researcher asked the participant to think aloud as they watched the video to 1) point out any accessibility issues related to the input test, and 2) point out any places where the test performed by the system could be improved. After each video, we interviewed each participant about how well the test in the video met their expectations, and how well the heuristics assisted them in finding any accessibility issues. 
Besides qualitative questions, we also asked the participants to provide 5-point Likert scale ratings on how similar the tests in the videos are to their manual tests, and how useful the heuristics are for tests to identify accessibility bugs.
Following the viewing of all three videos, we asked about the participants' overall attitude toward the system, how they envisioned incorporating it into their workflow, and any areas they identified for improvement. Additionally, we asked participants to provide 5-point Likert scale ratings assessing our system's usefulness in its current form and with ideal performance within their workflow.



\subsection{Participants}
\label{sec:user_study_participants}
We recruited 10 participants who are full-time employees at a large technology company. All participants perform manual accessibility tests as part of their professional work, having professional titles of accessibility QA testers and accessibility engineers.  We recruited participants via internal communication tools. In contrast to our formative study, all participants in this study were sighted and did not use screen readers. Two participants from our formative study, P5 and P6, also participated in this study. 
Since we did not collect information on the pronouns of our participants, we used the gender-neutral pronoun ``they/them'' to refer to all participants in our findings.
Interview questions and participant demographics are shared in Supplemental Materials.

\subsection{Data Collection and Analysis}

The data collected during the study includes audio and video recordings of the study sessions with the consent of the participants. We transcribed all the recordings into text format using an automated tool. The research team also took field notes during the session and used the notes to guide the analysis. The length of the sessions ranged from 29 minutes to 49 minutes, with an average length of 37 minutes. The interview with P9 only covered two videos (VO and BS) due to the participant's availability.

We performed a thematic analysis on the qualitative data from the user study \cite{guest2011applied}. Two authors of the paper first individually coded all the transcripts, then presented the codes to each other and collaboratively and iteratively constructed an affinity diagram of quotes and codes together to develop themes. The following findings section presents the resulting themes. We also reported the descriptive statistics of the data collected from the Likert scale rating questions, including the mean, standard deviation (SD) and sample size (N), to supplement our qualitative insights. 

\subsection{Findings}

\subsubsection{Performance of the Automatic Test Navigation} 
\paragraph{Automatic test navigation replicates manual test.}
Participants generally agreed that the system navigated applications in a similar path as they would conduct tests manually, especially in the BS and VO test cases.
For VO, Participants rated $4.60$ (SD = $0.52$, N = $10$) on average in the similarity regarding the navigation path between human testers and the AI (between ``very good match'' and ``extremely good match'' with their manual testing procedures). 
P3 was impressed by the system's ability to execute the test: ``my mind is blown that it was able to find that [shared button] buried within that actions menu.'' Similarly, in the BS test case, Participants rated $4.35$ (SD = $0.75$, N = $10$) on average. 
In P9's opinion, the system's heuristics might outperform most human testers in BS, since it could be subjective for a human tester to determine what consists of a button shape. 
Participants also reacted positively to the chapter feature, as it enabled efficient navigation through the video.

\paragraph{Differences in system and human approaches.}
Some of the approaches the system provided were different from what human testers would do. 
Compared to BS and VO, the system's performance in DT received $3.39$ (SD = $0.78$, N = $9$) on average, a relatively lower rating that was between ``moderately good match'' and ``good match'' with manual testing procedures. A main difference is that the system always relaunches the application between the tests of different text sizes, while human testers tend to use the control center to adjust text sizes within the application without relaunching it in order to mimic what a real user would do.
In fact, participants recognized a potential benefit of \systemname's approach, as it added an additional layer of testing: ``I really like that launches the app in between changing the text size, because I think it's a separate class of bug, whether or not, it responds to a change in text size versus having the text size there initially.'' (P8)
Similarly, P9 found in the VO example that the system waited for spoken output, which was not something that a human tester would typically do, but might be beneficial for more thorough tests. 

At the same time, participants also suggested that future versions of the system could enable exploratory and alternative navigation, as well as more in-depth tests of the UI structure. For example, for BS, participants mentioned that they would have explored more nested content in the application to ensure the Button Shape feature works for all elements (P2, P6).
For VO, participants wished the system could support alternative, non-linear pathways that VO users could go through (P7) and navigation using both swiping and tapping gestures (P4). 
Another common request is the ability to scroll through the screen of an application when testing display features like DT and BS. 

\paragraph{Reaction to navigation errors}
\label{sec:user_study.navigation_errors}
The VO video contains a slight error in the navigation: the navigation shares a \emph{show} instead of sharing an \emph{episode}. Only 2 out of 10 participants (P2 and P5) were able to identify this navigation error. Most participants ignored the error, potentially due to over-reliance on the automatic navigation, as P2 said, ``it worked well enough that I almost kind of let that slip. I needed to watch this video twice. Maybe I got over-reliant on [it].'' To address this error, P2 elaborated on how they would re-write the test instruction so that the agent could potentially correct the mistake: ``I would have [written], like, navigate to an episode, click the dot dot dot menu... 
I would suspect that this model would have done a better job finding the actual episode...'' P5, instead, described how they would navigate the application themselves based on the instruction: ``I would definitely do it the same route as it did through the more button, [but] instead of a certain episode, I would just switch it to show.''

\subsubsection{Identifying Accessibility Issues with Automatic Navigation} 

For all three cases of VO, BS, and DT, all participants spotted at least one accessibility issue, and agreed that the issues they discovered were significant enough to be filed in the internal bug reporting system within their company.

\paragraph{Heuristics aid discovery of issues.}
Overall, participants agreed that the heuristics provided by the system assisted them in finding the issues. For VO, BS, and DT respectively, participants on average rated $4.06$ (SD = $1.38$, N = $9$) (between ``useful'' and ``very useful''), $4.75$ (SD = $0.43$, N = $10$), and $3.67$ (SD = $1.09$, N = $9$) (between ``moderately useful'' and ``useful'') on the usefulness of the heuristics. Specifically, the potential issues flagged in the chapters allowed participants to navigate to where the issue was and review it with greater attention. The heuristics in particular helped direct testers' attention to the potential issues, which might otherwise be too subtle to discover: 
``Watching it in a video, as opposed to actually interacting with it, I think it is easier to potentially miss things... So, having some sort of automatic detection to surface things [is good].'' (P8) Even though they sometimes resulted in false positives, participants appreciated the heuristics providing an extra layer of caution, as P10 said,  ``I actively like the red [annotation boxes around potential issues] because I think the red is like `take a look at this' and then even if it's not necessarily an issue, that's not hurtful.''



\paragraph{Risks of over-reliance on heuristics.}
Participants expressed the concern of over-reliance on the heuristics provided by the system. In some sessions of our study, although participants found issues that were not marked by the heuristics, they were worried that those false negatives might bias testers: ``if things are marked as green, and maybe there actually is an issue in there, maybe that would dissuade somebody from looking there.'' (P10) 
This could influence testers of different experience levels differently. An experienced tester might rely on their expertise to find issues, while a novice tester might over-rely on the suggested bugs (or non-bugs) made by the system. As P8 explained: ``If somebody is kind of experienced with large text testing, they kind of know what to look for... If it's an inexperienced tester, they might not know that the false positives are false positives and might file bugs.'' (P8)

A mechanism to explain how the heuristics were generated and applied to the test cases might help with the issue of over-reliance. For example, P7 imagined it to be a series of ``human-readable strings, like what it actually found... human-readable descriptions of what the error is in addition to seeing the boxes.'' 
Other suggestions focus on making the heuristics more digestible for the testers. Currently, we show the heuristics as screenshots with annotations separate from the videos. Participants suggested it would be easier to comprehend the heuristics if they were encoded in the video and separated from regular chapters (P6), and only annotated the potential issues (P1). P7 brought up the idea to include a dashboard or summary mechanism in the system, so that a tester ``instead of just having a scrub through this video,'' could see ``a summary of the errors as well.'' 

\subsubsection{Integration in Accessibility Testing Workflow.}

Overall, participants reacted positively to our system. Participants rated $4.70$ (SD = $0.48$, N = $10$) (between ``useful'' and ``very useful'') on average for how useful the system is in their existing workflow if it performs extremely well, and $3.95$ (SD = $0.96$, N = $10$) (between ``moderately useful'' and ``useful'') on average to the system in its current form. Participants expressed excitement about the potential of integrating the system and bringing automation to their workflow. For instance, when asked for a rating on the overall usefulness of the system, P3 answered: ``[I will rate] it like a 5 million... Even with the current limitations, it is very useful... just being able to feed it some real simple steps and have it do anything at all is massively powerful.'' The next sections unpack a range of ways that \systemname might be integrated into existing test workflows. 

\paragraph{Automating test planning.}

A compelling use case for \systemname is to automate the planning and setup of the test, which, according to our participants, is a time-consuming part of accessibility testing as it can involve an excessive amount of manual work to ``go through and find all of the labels to tap through'' (P3).
The step-by-step executable test plan generated from natural language from our system can reduce the amount of tedious work: ``rather than having to hard code navigation logic, it seems that this is able to determine those pathways for you... I think this idea is really awesome and would definitely save a lot of hours of not having to hard code the setup steps to go through a workflow with VoiceOver.'' (P4) P4 also envisioned using the system as a test authoring tool, which can generate templates that can be run daily.


\paragraph{Complementing manual tests.}
\label{sec:user_study.running_test}
Participants found the system helpful in reducing workload and saving time in running tests. Some participants would like to embrace the automation provided by the system, keeping the system running a large scale of tests in the background while the team could focus on more important tasks: 
``you can run it in an automated fashion. You don't need to be there. You can run it overnight. You can run it continually without scaling up some more people'' (P7).
As P8 imagined, ``this could run on each new build [of the software], and then what all the QA engineer has to do is potentially a review about an hour's worth of videos that were generated by the system, potentially automatically flagging issues.'' The system can also provide consistency and standardization in tests, which ``ensure[s] that everything is run the same way every time.'' (P8)



At the same time, some participants are more cautious about automation and would like to use the system as a supplement to their manual work. P4 believed that even with the flagged issues, they would still pay attention to the system-generated videos to a degree similar to how they would test them manually. P1 imagined that they would still test manually, but would use the video as validation of their tests ``to see if it could catch things that I couldn't catch.'' (P1) Some also imagined handing lower-risk tests, such as testing Button Shapes, to the system, while using the time saved by the system to manually and carefully test higher-risk tests that will be a regulatory blocker. (P2)



\paragraph{Aiding downstream bug reporting.}
The videos generated by the system can also facilitate bug reporting in the downstream pipeline. Participants agreed that the video along with the chapters generated by the system could be used to triage any accessibility issues that they would report to the engineering teams. In their current practice, testers would sometimes include screenshots or screen recording video clips to demonstrate the discovered issue. Our system prepared a navigable video automatically, streamlining this process: ``I thought to be able to jump to specifically when the issue is and scrub a couple of seconds back or a couple seconds forward is super useful for engineering.'' (P7) 




\paragraph{Educating novices about accessibility testing.}
The system can also serve as an educational tool for those who are new to accessibility tests. The system can not only help new QA professionals, but also developers from under-resourced teams where there are no dedicated QA teams or pipelines. For example, P2 found the videos and heuristics helpful in terms of demonstrating certain accessibility bugs that people should be looking for: ``This will be very useful for some of the folks that never do accessibility testing and [for] they [to] have a context or starting point for even knowing what a VoiceOver bug is.'' (P2) In a way, our system has the potential to demonstrate and raise awareness of accessibility issues among broader developer communities, even for those who do not have QA resources. 



\section{Discussion}
Accessibility QA testing is still by-and-large a manual effort and there are benefits to not leaving such testing up to full automation~\cite{mankoff2005your}. The majority of QA testers we interviewed desired more automation to free up time for more complex testing. However, they lack the time and resources to effectively use existing automation methods. With \systemname, a key goal is to use testers' existing metadata (e.g., databases of manual instructions) and build a tool to complement existing workflows. Our user study indicates that \systemname, even in its current form, can be useful in their workflows. \systemname also serves as an initial exploration into using recent advances in LLMs and UI navigation in accessibility testing workflows, which other systems can build upon. In this section, we discuss some limitations of our evaluation and the \systemname system that we plan to address in future work, and potential extensions of \systemname beyond accessibility testing workflows.  


\subsection{Differences between automated navigation and manual testing}
\systemname employs one workflow specifically for VoiceOver tests, where the system uses forward swipes until finding a target element before activating it. As shown in the user study, this may not reflect how a VoiceOver user might navigate the task as the user may have prior knowledge of the app structure. This would enable users to skip around to various parts of the screen to activate the desired element. While sometimes such differences can be complementary test strategies, future versions of the system could explore how to simulate alternative patterns of interactions. 


\subsection{Improving navigation performance}
While \systemname achieves reasonable test replay accuracy, it can encounter errors arising from a lack of sufficient knowledge about apps or understanding when to stop (see \autoref{section:technical_evaluation}).
We expect that improvements in modeling (i.e., by fine-tuning a model on successful navigation paths or integrating existing app knowledge into prompts~\cite{wen2023empowering}) can improve navigation performance in future versions of \systemname. Other approaches, such as using multimodal models~\cite{jiang2023iluvui}, could be considered for future iterations.

\subsection{Mitigating errors and over-reliance}
Like all machine learning and heuristic-based systems, \systemname is not expected to always produce perfect output. However, it is important to mitigate the risk of these errors on QA testers.
Prior works have shown there is a risk of over-reliance on AI systems since users can view the AI as an authority and be reluctant to challenge it~\cite{Bhat2021HowDP, chen2023understanding}. This is also the case for the navigation and heuristics of \systemname.
For example, only 2 out of 10 user study participants were able to spot the navigation error in the VoiceOver example (see \autoref{sec:user_study.navigation_errors}). 
While evaluating the correctness of LLM-based systems remains an active area of research~\cite{anthropicEval2022}, there are additional techniques that could be considered for future work to enable \systemname to report whether it executed a navigation task correctly. For example, the navigation path itself could be evaluated through heuristics, another LLM, or by using existing knowledge of apps.
Another way to mitigate over-reliance in future work would be to provide transparency signals, such as confidence scores and textual explanations of how the predictions were made, echoing design guidelines on transparency and explainability for human-AI collaboration \cite{amershi2019guidelines}.

\subsection{Limitations in the User study}
Our user study had participants watch and comment on videos generated by \systemname. Our study design mimicked how accessibility testers would interact with \systemname in their actual workflows (i.e., reviewing videos generated by an automatic system and spotting accessibility issues, as elaborated in \autoref{sec:user_study.running_test}), but this design has some limitations. First, we only showed the same set of 3 videos to all the participants. Although the set of videos covers different types of accessibility tests, participants' feedback could be biased by this limited set of examples. Second, we only showed users videos where navigation mostly worked to probe how they would use the system in their workflow. We did not show examples where the replay failed, and therefore were not able to collect user feedback on failed replay and how it would be handled. 
Third, in order to keep user study sessions short, the participants did not directly write their own tests and generate videos using the tool themselves. In future work, we plan to deploy \systemname in a longitudinal study so that we can better understand how QA testers instruct the system and interact with its output.

\subsection{Accessibility of \systemname}
One key limitation of \systemname currently is its output video format which is not by default accessible to screen reader users. People with disabilities are commonly employed in accessibility testing such as non-sighted screen reader testers. \systemname should make the video format accessible by ensuring all visual content is described -- such as heuristic boxes, screen changes, and chapter annotations. Non-sighted users may also find other output formats more useful. The screen reader user in our formative study requested \systemname replay test cases live on a local device to enable them to take control, which is feasible and something we plan to do in future work. Lastly, future versions of \systemname should be accessible to testers beyond screen reader use cases (e.g., testers with motor impairments).  

\subsection{Accessibility feature support and generalizability}
Our studies uncovered the need to support testing additional accessibility features beyond the four that \systemname supports. Future versions of \systemname can support more navigational accessibility services (e.g., Voice Control) and other accessibility settings (e.g., display features such as contrast adjustment and motion reduction) provided the device's operating system provides APIs to control those features. 
\systemname currently surfaces some potential accessibility issues through its heuristics (e.g., Dynamic Type resizing issues); however, these do not cover all accessibility issues we could surface. Future versions of \systemname could incorporate existing accessibility inspection tools similar to Groundhog~\cite{salehnamadi2022groundhog} to report issues such as missing UI element descriptions or minimum target sizes. We could also add a dashboard to summarize the issues found during \systemname's replay, as study participants proposed. \systemname could also consider focused testing for specific accessibility needs. For example, if a test is for users with motor impairments, issues like target size would be important to surface.

Lastly, we have only built \systemname to work with the iOS operating system. However, the system architecture and workflow should be extensible to other platforms where provided APIs are available to control the accessibility features under test. A body of work has explored general UI navigation in other platforms~\cite{feng2023prompting, yangMappingGrounded, UGIF, wen2023empowering}. 

\subsection{More applications of the \systemname system}
We have so far evaluated \systemname for QA testing, but there are many opportunities beyond this as indicated by our user study and other work in this area. One that we would like to explore is using this system as a tool to help novice developers better understand the behaviors of accessibility features and how they should be tested by generating realistic simulations of behavior on their own apps. Additionally, natural language instructions are used in manual UI testing, bug reports, and reproduction steps~\cite{feng2023prompting}, and natural language automation systems may benefit from the techniques we present in this paper to reconstruct these types of tests. These are examples of use cases we hope to explore in future work.

\section{Conclusion}
In this paper, we presented a system to support accessibility test interpretation and replay through natural language instructions. Our system achieves good technical success in replaying realistic manual test instructions, achieving 70\% and 85\% navigation replay success. We evaluated our system with 10 professional accessibility testers who would find the system very useful in their work and revealed a number of promising future opportunities and insights into how we can leverage LLM-based task automation within accessibility testing. 

\bibliographystyle{ACM-Reference-Format}
\bibliography{references}

\end{document}